\documentstyle[epsfig, twocolumn]{esa_sp_latex}

\def\grs{\mbox{GRS } 1915+105}
\def\ss{\mbox{SS } 433}
\def\cygtrois{\mbox{Cyg X-} 3}
\def\cygun{\mbox{Cyg X-} 1}

\def\plusoumoins{\, \pm \,}

\def\degr{^{o}}
\def\ergs{\mbox{ erg s}^{-1}}
\def\kpc{\mbox{ kpc}}
\def\mags{\mbox{ magnitudes}}
\def\micrometre{\mbox{ } \mu \mbox{m}}
\def\Av{A_{\rm v}}

\def\K{\mbox{ K}}

\begin{document}
%

\parindent 0pt
\parskip 10pt plus 1pt minus 1pt
\hoffset=-1.5truecm
\topmargin=-1.0cm
\textwidth 17.1truecm \columnsep 1truecm \columnseprule 0pt 

\title{\bf INFRARED OBSERVATIONS AND ENERGETIC OUTBURST OF GRS 1915+105}

\author{{\bf S.~Chaty$^{1}$, I.F.~Mirabel$^{1,2}$, L.F.~Rodr\'{\i}guez$^{3}$, P.-A.~Duc$^{4}$, M.~Sauvage$^{1}$, A.J.~Castro-Tirado$^{5}$, P.~Callanan$^{6}$} \vspace{2mm} \\
$^{1}$Service d'Astrophysique, CEA/DSM/DAPNIA/SAp, Centre d'Etudes de Saclay, \\
F-91191 Gif-sur-Yvette Cedex, France \\
$^{2}$Instituto de Astronom\'{\i}a y F\'{\i}sica del Espacio, Argentina \\
$^{3}$Instituto de Astronom\'{\i}a, UNAM, Apdo. Postal 70-264, 04510 México, D.F., Mexico \\
$^{4}$European Southern Observatory, Karl-Schwarzschild-Strasse 2, 85748 Garching bei M\"unchen, Germany \\
$^{5}$Laboratorio de Astrof\'{\i}sica Espacial y F\'{\i}sica Espacial, INTA, P.O. Box 50727, E-28080 Madrid, Spain \\
$^{6}$Smithsonian Astrophysics Observatory, 60 Garden Street, Cambridge, MA 02138, USA}

\maketitle

\begin{abstract}

Multiple near-infrared wavelengths observations, carried out since 1993 on the galactic superluminal source of relativistic ejections $\grs$, have yielded three important results. \\
1) The infrared counterpart of $\grs$ exhibits various variations in the $1.2 \, \mu m$ -- $2.2 \, \mu m$ band: the strongest are of $\sim$ 1 magnitude in a few hours and of $\sim$ 2 magnitudes over longer intervals of time (\cite{cha96}). \\
2) The infrared properties of $\grs$ are strikingly similar to those of $\ss$, and unlike those of any other known stellar source in the Galaxy. The absolute magnitudes, colors, and time variabilities of these two sources of relativistic ejections suggest that $\grs$, like $\ss$, consists of a collapsed object (neutron star or black hole) with a thick accretion disk in a high-mass-luminous binary system (\cite{cha96}). \\
3) During an intense and long-term X-ray outburst of $\grs$ in 1995 August, where a pair of radio-emitting clouds emerged from the compact core in opposite directions at relativistic speeds, we observed the time-delayed reverberation of this radio flare/ejection event in the infrared wavelengths. The observed spectrum of the enhanced infrared emission suggests the appearance of a warm dust component (\cite{mir96a}).\vspace {5pt} \\


\end{abstract}

\section{INTRODUCTION}

The hard X-ray transient $\grs$ was discovered in the constellation of Aquila, on 15 August 1992, by the WATCH all-sky X-ray monitor on board GRANAT (\cite{cas92}). Frequently radiating $\sim 3 \times 10^{38} \ergs$ in the X-rays, at the kinematic distance of $D = 12.5 \plusoumoins 1.5 \kpc$ from the Sun, $\grs$ is one of the most luminous X-ray sources in the Galaxy. Since this source has a fairly hard spectrum with emission up to 220 keV, and a variable spectral index between $-2$ and $-2.8$, observed by BATSE on the Compton Gamma-Ray Observatory (GRO), $\grs$ is likely to be a collapsed object, perhaps a black hole in a binary system (\cite{har94}). Thanks to the arcmin location by SIGMA on GRANAT (\cite{fin94}), \cite*{mir94b} discovered its radio counterpart. After the detection of relativistic ejections of plasma clouds with apparent superluminal motions, $\grs$ was called the galactic superluminal source (\cite{mir94a}; \cite{mir95}; \cite{rod95}).

Following a multiwavelength study of this source, \cite*{mir94b} looked for an optical counterpart, but without success, due to interstellar extinction, since $\grs$ is located near the galactic plane at $\mbox{\it l} = 45.37 \degr$, $\mbox{\it b} = -0.22 \degr$. Nevertheless, the counterpart of $\grs$ was discovered at near-infrared wavelengths (\cite{mir94b}). Since its detection in June 1993, this infrared counterpart has often been observed in the near-infrared wavelengths, and we report here these multiple observations, trying to determine the light curve of this source over long and short time scales. These observations also allow us to determine the spectrum of the source, and to see how it changes with the luminosity. These studies help to constrain the nature of the source, and also to understand what its environment is, perhaps composed of heated dust. This can be determined by infrared observations, taken during an intense and long-term X-ray outburst in 1995 August, showing the interactions between $\grs$ and its environment. \\

\section{THE OBSERVATIONS OF THE INFRARED COUNTERPART OF $\grs$}

\cite*{mir94b} showed that there is no visual counterpart of $\grs$ brighter than $R = 21 \mags$. Using the NTT with a Gunn-z filter on 9 July 1994, we observed a faint counterpart at $\sim 1 \micrometre$, consistent with the $I = 23.4 \mags$ counterpart reported by \cite*{boe96}. We carried out infrared observations of $\grs$ at the European Southern Observatory (ESO) with the ESO/MPI 2.2~m telescope on 4 and 5 June 1993 and from 5 to 8 July 1994 with the IRAC2(b) camera (\cite{mir94b}), in the J ($1.25 \micrometre$), H ($1.65 \micrometre$) and K ($2.2 \micrometre$) bands. The typical seeing for these observations was $1.2 \mbox{ arcsec}$. Follow-up observations were performed at our request by S. Massey at the 3.6~m Canada-France-Hawaii Telescope (hereafter CFHT) on Mauna Kea on 16 August 1994, with the Redeye camera, with a typical seeing of $0.6 \mbox{ arcsec}$. The data reduction methods of these images are described in \cite*{cha96}. The J, H and K-magnitudes are given in Table \ref{varmag}.

\begin{figure}
\centerline{\psfig{file=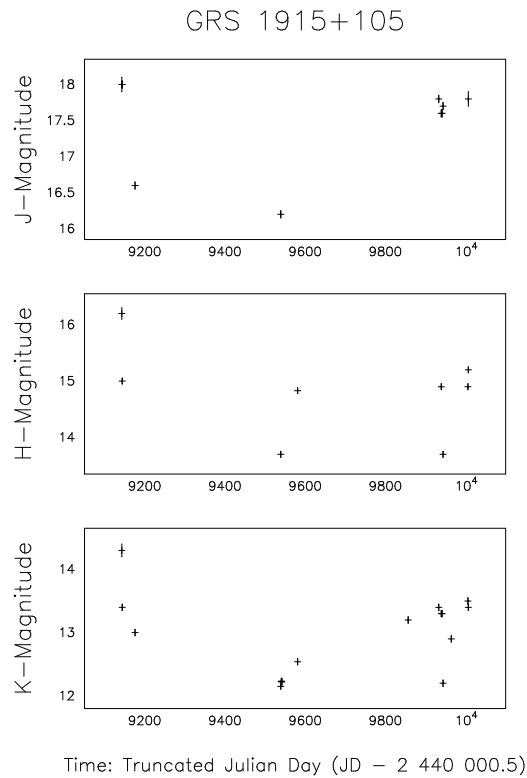,angle=-90,width=14.cm}} 
\caption[]{\label{GRSvarie} Time variation of J ($1.25 \micrometre$), H ($1.65 \micrometre$) and K-band ($2.2 \micrometre$) luminosities of the source $\grs$ from 4 June 1993 to 17 October 1995.}
\end{figure}

The variations in the J, H and K-bands, reported in Fig. \ref{GRSvarie}, show that $\grs$ exhibits strong short-term variability in the J, H and K-bands, in intervals of less than 24 hours as well as strong long-term variability over intervals from one month to one year. Indeed, the luminosity of $\grs$ increased by nearly 1 magnitude in H and K between the nights of 4 and 5 June 1993, and, between 4 June 1993 and 5 July 1994, there was a change of nearly $2 \mags$ in J, $2.5 \mags$ in H, and $2.1 \mags$ in K. From Table \ref{varmag} it seems that the infrared colors change with luminosity. The rapid increase of 1 magnitude observed in an interval of 24 hours in June 1993 could result from occultation. It is also interesting to note that this rapid variation of the infrared luminosity occurred in a period when the source was strong and showing rapid variations of luminosity in the 8--60 keV energy band observed by WATCH (\cite{saz94}), and in the 20--100 keV energy band observed by BATSE (\cite{har94}).

\begin{table*}
\caption[]{\label{varmag} Infrared magnitudes of GRS 1915+105.}
\begin{flushleft}
\begin{tabular}{cccccccc} 
\hline \noalign{\smallskip}
{\bf Date} & {\bf UT} & {\bf TJD$^{1}$} & {\bf J} & {\bf H} & {\bf K} & {\bf Telescope} & {\bf Reference} \\
 & (hs) & & ($1.25 \micrometre$) & ($1.65 \micrometre$) & ($2.2 \micrometre$) & & \\
\noalign{\smallskip}
\hline \noalign{\smallskip}
04/06/1993 & 1.9 & 9143.1 & $\geq 18 \plusoumoins 0.2$ & $16.2 \plusoumoins 0.2$ & $14.3 \plusoumoins 0.2$ & ESO 2.2m & \\
05/06/1993 & 5.0 & 9144.2 & $18 \plusoumoins 0.1$ & $15.0 \plusoumoins 0.1$ & $13.4 \plusoumoins 0.1$ & ESO 2.2m & \\
07/07/1993 & & 9176 & $16.6 \plusoumoins 0.1$ & & $13.0 \plusoumoins 0.1$ & UKIRT 3.8 m & 1 \\
05/07/1994 & 7.6 & 9539.3 & $16.2 \plusoumoins 0.1$ & $13.7 \plusoumoins 0.1$ & $12.15 \plusoumoins 0.08$ & ESO 2.2m & \\
06/07/1994 & 7.5 & 9540.3 & & & $12.23 \plusoumoins 0.04$ & ESO 2.2m & \\
07/07/1994 & 7.5 & 9541.3 & & & $12.23 \plusoumoins 0.1$ & ESO 2.2m & \\
08/07/1994 & 5.7 & 9542.2 & & & $12.22 \plusoumoins 0.1$ & ESO 2.2m & \\
16/08/1994 & 9.6 & 9581.4 & & $14.83 \plusoumoins 0.1$ & $12.54 \plusoumoins 0.1$ & CFHT 3.6 m & \\
19/05/1995 & & 9857 & & & $13.2 \plusoumoins 0.1$ & UKIRT 3.8 m & 2 \\
04/08/1995 & 5.0 & 9933.2 & $17.8 \plusoumoins 0.1$ & & $13.4 \plusoumoins 0.1$ & ESO 2.2 m & \\
10/08/1995 & 9.5 & 9939.4 & $17.6 \plusoumoins 0.1$ & $14.9 \plusoumoins 0.1$ & $13.3 \plusoumoins 0.1$ & UKIRT 3.8 m & \\
12/08/1995 & 9.4 & 9941.4 & $17.6 \plusoumoins 0.1$ & & $13.3 \plusoumoins 0.1$ & Lick 3 m & \\
15/08/1995 & 0.0 & 9944.0 & $17.7 \plusoumoins 0.1$ & $13.7 \plusoumoins 0.1$ & $12.2 \plusoumoins 0.1$ & ESO 2.2 m & \\
04/09/1995 & 8.7 & 9964.4 & & & $\geq$ $12.9 \plusoumoins 0.1$ & UKIRT 3.8 m & 3 \\
16/10/1995 & 0.1 & 10006.0 & & $14.9 \plusoumoins 0.1$ & $13.5 \plusoumoins 0.1$ & Kitt Peak 2.1 m & 4 \\
17/10/1995 & 0.1 & 10007.0 & $17.8 \plusoumoins 0.2$ & $15.2 \plusoumoins 0.1$ & $13.4 \plusoumoins 0.1$ & Kitt Peak 2.1 m & 4 \\
\noalign{\smallskip}
\hline
\end{tabular}
\end{flushleft}
{\bf References:} \\
1: From \cite*{cas93} \\
2: From \cite*{geb95} \\
3: From \cite*{mir96b} \\
4: From \cite*{eik95} \\
$^{1}$TJD: Truncated Julian Date (JD - 2 440 000.5)
\end{table*}
%

\section{THE LIKELY NATURE OF $\grs$, DERIVED FROM THE INFRARED OBSERVATIONS}

From the apparent magnitudes in Table \ref{varmag} we derived the absolute magnitudes, corrected for interstellar extinction, using a visual absorption of $\Av = 26.5 \plusoumoins 1\mags$, and the kinematic distance $D = 12.5 \plusoumoins 1.5 \kpc$. The absorptions in the J, H and K bands are $A_{\rm J} = 7.1 \plusoumoins 0.2$, $A_{\rm H} = 4.1 \plusoumoins 0.2$ and $A_{\rm K} = 3.0 \plusoumoins 0.1 \mags$ respectively. The infrared emission of $\grs$ cannot arise {\bf only} in the photosphere of the secondary star: 1) because of the shape of the spectrum, which cannot be reproduced by photospheric emission from any stellar type (e.g. \cite{koo83}), and 2) because of the rapid variations in luminosity and energy distribution (see Table \ref{varmag}). Therefore, besides the photospheric emission from the secondary, there must be an additional source of infrared emission in $\grs$.

\begin{figure}
\centerline{\psfig{file=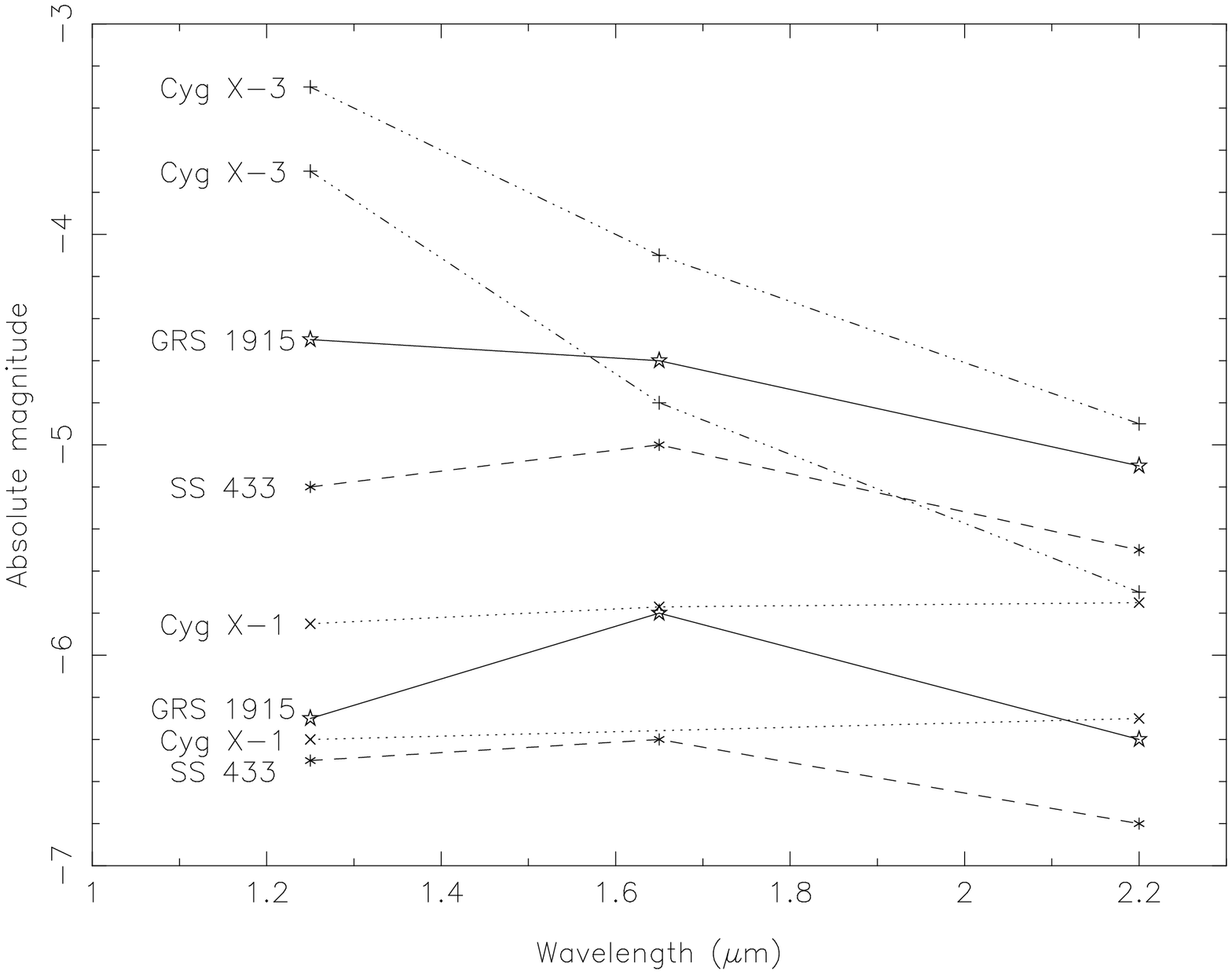,angle=-90.,width=12.cm}}  
\caption[]{\label{grsm143} Infrared energy distributions of $\grs$, $\ss$, $\cygun$ and $\cygtrois$ for the periods of minimum and maximum luminosity.}
\end{figure}

The energy distributions of the most well studied galactic X-ray sources are shown in Fig. \ref{grsm143}. To derive the absolute magnitudes of $\ss$ we assumed the kinematic distance of $4.2 \plusoumoins 0.5 \kpc$ (\cite{gor82}) and a visual absorption $\Av = 7.25 \plusoumoins 0.25 \mags$ (\cite{ala80}). The estimated errors of the absolute magnitudes of the X-ray sources take into account the uncertainties on the distance and interstellar absorption. Besides the time variations, the infrared absolute magnitudes and colors of $\grs$ are strikingly similar to the classic source of relativistic jets $\ss$. This similarity in the observed infrared properties suggests that $\ss$ and $\grs$ are similar systems. The infrared emission of $\ss$ arises in a high-mass binary of type late O or early B or Be, with possible contributions of free-free emission from an ionized plasma at $T \sim 7\,500 \K$, an accretion disk, and/or even the jets (e.g. \cite{mar84}). Within the context of a binary model with an accretion disk, \cite*{kod85} conclude that the observed infrared flux in the $\ss$ system comes mostly from an accretion disk around the compact object of the binary system, and that the day-to-day variations may be due to different configurations of disk structures, depending on the mass supply and the internal magnetohydrodynamic balances. Therefore, by analogy with $\ss$, $\grs$ could be a collapsed object with a thick accretion disk in a hot and luminous high-mass binary.

\section{INFRARED OBSERVATIONS OF AN ENERGETIC OUTBURST IN $\grs$}

\begin{figure}
\centerline{\psfig{file=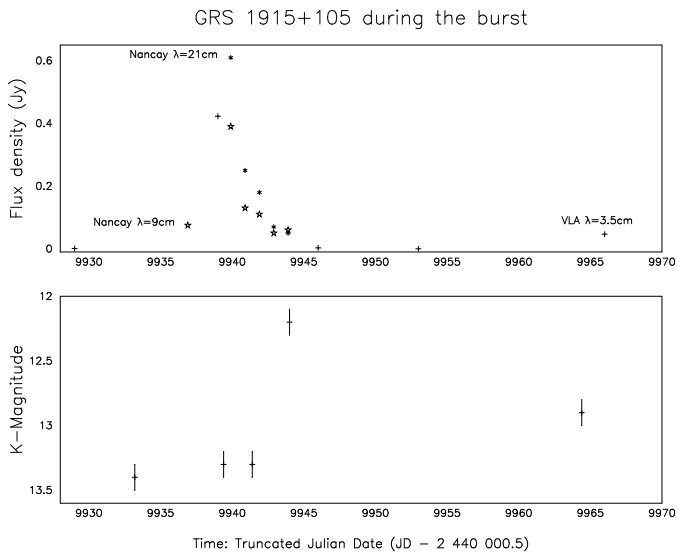,angle=-90.,width=10.cm}} 
\caption[]{\label{grsburst} Top: Radio observations of $\grs$ around the 1995 August outburst/ejection event as observed with the VLA at $\lambda$ = 3.5 cm, and with the Nançay radiotelescope at $\lambda$ = 9 cm and $\lambda$ = 20 cm. Bottom: Infrared K ($2.2 \micrometre$) magnitudes of $\grs$. Note the time delay of the infrared brightening relative to the time of peak radio emission.}
\end{figure}

During an intense X-ray and radio outburst of $\grs$ in 1995 August (\cite{saz96}, \cite{har95}, \cite{fos96}), we observed a pair of bright radio-emitting clouds with the Very Large Array, that we were able to follow for an interval of three weeks, emerging from the compact radio core in opposite directions, and at relativistic speeds. In the course of a multiwavelength study, we observed $\grs$ in this period, at near-infrared wavelengths, in the J(1.25 $\micrometre$), H(1.65 $\micrometre$) and K(2.2 $\micrometre$) bands, with various infrared facilities, to study the long term variability of the infrared stellar counterpart of $\grs$ (\cite{cha96}), and the correlation between the infrared and radio/X-ray emission. We discovered an infrared outburst (\cite{mir96a}), reported in Fig. \ref{grsburst}. Due to the time-delayed reverberation of this sudden and major radio flare/ejection event, the infrared outburst was detected between two and five days after the radio outburst. Therefore, the cause of the infrared response to this impulsive event must be at $\geq 2$ light-days from the compact object.

The source became redder by J-K $= 1.2 \mags$, and brightened by $\sim$ 1 magnitude in K (2.2 $\micrometre$). The $1.0-2.5 \micrometre$ continuum rising to the red suggests the appearance of a warm dust emitting component, and thermal reradiation from heated dust, since there is a cutoff of the enhanced radiation below $\sim 1.45 \micrometre$. Furthermore, a trend towards redder colors as the star becomes brighter has also been observed in $\ss$ (\cite{cat81}). The infrared lag due to the light crossing time, and the enhancement of the infrared luminosity emitted in the H and K bands allow us to infer the mass of warm dust surrounding $\grs$, and its distance from $\grs$. The thermal energy reradiated from heated dust in the near-infrared was $\sim 10$ \% of the X-ray luminosity of the source, and amounts to about $0.1$ \% of the typical kinetic energy in the bulk motion of the relativistic ejecta in $\grs$. At a distance of 500 a.u. (3 light-days) and with an X-ray luminosity of $10^{37} \ergs$ we expect an equilibrium dust temperature of only $\sim 100$ K. We then believe that the near-infrared emission observed could be coming from small grains out of equilibrium with the X-ray field, and that most of the dust is radiating at lower temperatures. Sensitive IR observations at longer wavelengths will be carried out with ISO to test this hypothesis.

\section{CONCLUSION}

Thanks to many near-infrared wavelengths observations, carried out since 1993 on the galactic superluminal source of relativistic ejections $\grs$, we derived three important results, that give a better understanding of its nature, and lead to a severe constraint on its nature.

We have shown that the infrared counterpart of $\grs$ exhibits short-term variability in intervals of less than 24 hours as well as long-term variability over intervals from one month to one year.
We deduced from these infrared observations the fact that the infrared emission of $\grs$ could not arise {\bf only} in the photosphere of the secondary star, and that the infrared absolute magnitudes and colors of $\grs$ were very similar to those of the classic source of relativistic jets $\ss$, and unlike those of other well known galactic X-ray sources during minimum and maximum luminosity. Therefore, $\grs$ could be a collapsed object (neutron star or black hole) with a thick accretion disk in a high-mass-luminous binary system.
In 1995 August, when in the source GRS 1915+105 was undergoing an intense and long-term X-ray outburst, a pair of radio-emitting clouds emerged from the compact core in opposite directions, at relativistic speeds. In the infrared wavelengths, we observed the time-delayed reverberation of this radio flare/ejection event. Thanks to the observed spectrum of the enhanced infrared emission, we showed that the appearance of a warm dust component was suggested.

\clearpage


\end{document}